# An Efficient Read Dominant Data Replication Protocol under Serial Isolation using Quorum Consensus Approach

Vinit Kumar[1] and Ajay Agarwal[2]

[1] Associate Professor with the Krishna Engineering College, Ghaziabad, India.
(Phone: +919971087809; e-mail: vinitbaghel@gmail.com)

[2] Professor with SRM University, DELHI-NCR Campus, Modinagar, Ghaziabad, India.
(Phone: +919917083437; e-mail: ajay.aagar@gmail.com)

Abstract

In distributed systems, data replication provides better availability, higher read capacity, improved access efficiency and lower bandwidth requirements in the system. In this paper, we propose a significantly efficient approach of the data replication for serial isolation by using newly proposed Circular quorum systems. This paper has three major contributions. First, we have proposed the Circular quorum systems that generalize the various existing quorum systems, such as Read-one-write-all (*ROWA*) quorum systems, Majority quorum systems, Grid quorum systems, Diamond quorum systems, D-Space quorum systems, Multi-dimensional-grid quorum systems and Generalized-grid quorum systems. Second, Circular quorum systems not only generalizes but also improves the performance over existing quorum systems of their category. Third, we proposed a highly available Circular quorum consensus protocol for data replication under serial isolation level that uses a suitable Circular quorum system for read dominant scenario.

## 1. Introduction

Quorum systems are the basic tools that provide coordination among multiple concurrent processes in various distributed applications. Quorum systems are attractive because they provide a decentralized approach of coordination that tolerates failures. Quorum systems are also interesting for large-scale systems, because it is possible to design a quorum system such that the quorum size increases much slower than the system size. Therefore, it is possible to provide a very high availability, fault tolerance and reasonable communication cost. In addition, Quorum systems are considered to be the best for serial isolation level or one-copy-serializability [1] [2] (i.e. each transaction accesses the latest committed database state) even in the case of network partitions. Furthermore, these are also useful for reducing and balancing the system load.

*Quorum system* in data replication is a collection of the set of read quorums and the set of write quorums. Whereas, any read quorum intersects with any write quorum. As well as, any write quorum intersects with any other write quorum.

We can classify the *Quorum systems* based on intersection properties as *strict quorum systems* and *probabilistic quorum systems*. Moreover, based on site failures, we can classify the quorum systems as *crash-stop quorum systems* and *byzantine quorum systems*. Furthermore, based on sites leave or join the system at runtime, we can classify the quorum systems as *static quorum systems* and *dynamic quorum systems*. This paper contributes in the direction of *crash stop, static and strict quorum systems*.

We can classify the data replication as *update anywhere replication* and *primary copy replication* (also called *Active replication* and *Passive replication*). Under update anywhere replication, a client may approach to any of the replica server for either update or query transaction. Under primary copy replication, (also called primary/backup or coordinator/cohort replication) a client approaches to a primary replica only for update or query transaction, moreover it may also approach to other backup replicas for query transaction. This paper contributes in the direction of active replication.

We can also classify data replication as *Synchronous replication* and *Asynchronous replication* (also called *Eager* or *Pessimistic replication* and *Lazy* or *Optimistic replication*). *Synchronous replication:* clients consider the update operation as complete only when both local and remote replicas acknowledge the completion of the operation. It guarantees no data loss but client has to wait until the completion of the operation at all replicas.

*Asynchronous replication:* clients consider the update operation as complete as soon as local replica acknowledges it. Performance is greatly increased, but in case the failure of a local replica, the remote replica may not have the current copy of data and most recent data may be lost. This paper contributes in the direction of Synchronous replication.

We may further classify data replication as *State-machine replication* and *Non-state-machine replication.* State-machine replication ensures that all replica servers that start with same consistent state will remain consistent as long as they apply the same deterministic actions in the same order. In addition, all the replica servers attain the same consistent state after finishing a transaction. Under a Non-state-machine replication, only a sub set of the total replicas (not all) attain the same consistent state after the finishing an update transaction. This paper contributes largely in the direction of Non-state-machine replication but in one special case (*Read-one-write-all*), our approach also works as a state-machine.

*Read-one-write-all* (ROWA) *protocol* [3] is the one of the simplest protocol for data replication that implements the state machine replication. Whereas, read operations read any copy, while write operations write all the copies. This protocol is very much attractive because of the least read quorum size, optimal read availability and optimal read capacity. Nevertheless, this is at the cost of the least write availability and no fault tolerance. All copies need to be operational for a write to proceed.

*Available copies protocol* [4] [5] enhances the ROWA approach via increasing the write availability and fault tolerance. Any read operation requires only one copy of the data item while any write operation requires all the available copies of that data item. This approach handles site failures but does not support communication failures. Therefore, approach may not attain consistency under network partitions.

*Virtual partition protocol* [6] [7]may enhance the ROWA by means of higher write availability, higher fault tolerance and consistency under network partitions. This protocol build the virtual partitions among all the nodes such that every node belongs to only one partition and all nodes in a partition believe that they can always communicate to each other. If any node finds that the communication among all nodes in their partition is not possible than again virtual partition building procedure starts. For any read or write operation any virtual partition must contain the majority of active node. In addition, read operation needs to access only one node while write operation needs to write all nodes of their partition. Moreover, creation of the virtual partitions takes some extra effort while in a larger system where any failure will lead to rebuild the virtual partitions that one is a costly affair. Furthermore, system may only work up to when there is a partition where the majority nodes are there that can communicate to each other.

*Viewstamp replication protocol* [8], this one is a simplification and modification of the aforementioned virtual partition protocol. The same concept of virtual partitioning is applied here but unlike virtual partitioning protocol, it works on passive replication model.

*Commit protocols* such as Two-phase commit (2PC) protocol [9] and Three-phase commit (3PC) protocol [10] are one of the fundamental techniques used to provide a consistent view in a distributed replicated database system and works on a passive and state-machine replication model. The 2PC protocol cannot tolerate any fault while 3PC tolerates a single fault in the system by seeing their local state only. In most of the applications, this much of fault tolerance is not appropriate.

*Virtual Synchrony approach* [11] as suggested by Kenneth P. Birman is a quite promising solution to implement the state machine replication. Virtual synchrony systems allow processes running in a network to organize themselves into *process groups*, and to send messages to groups (Group Communication). Each message is delivered either to all the group members or none of them (Atomic broadcast), in the identical order (Message ordering). Every group member sees the same events (group membership changes, and incoming messages) and in the same order. The system tracks the group members, and each time, it informs the changes in membership to all members. We call such an event as *view change*. It increases the fault tolerance in the system. Many Researchers have shown their interest in this approach. However, network partitioning is a major problem for such systems.

*Consensus approach* [12]-[17] as suggested by Leslie Lamport to implement the state machine replication is also a very promising approach and many researchers have shown their interest in this approach. Under this approach, any process can propose their value to the collection of processes. In addition, consensus algorithm chooses a single value from all the proposed values. If there is a chosen value, then all non-faulty processes should learn that chosen value. We can use consensus approach for both active and passive replication model. Leslie Lamport has given the name to such replication approaches as *Paxos* that improves the commit protocols in order to enhance the fault tolerance. *Paxos commit* protocols such as Paxos, Cheap Paxos, Generalized Paxos, Fast Paxos and Byzantine Paxos etc.

*Quorum consensus approach* [18]-[39], implements the non-state-machine replication model, under this model we require only a subset (quorum) of the total replica servers for execution of the query or update operations. In quorum consensus protocols, a read operation requires to read all the copies of the read quorum and select the latest version among them. While, a write operation requires to write all copies in the write quorum with a new increased version number. A write quorum intersects with each other write quorum and each read quorum, therefore at any time only one write operation is possible even in concurrent scenario. This approach may tolerate the various types of failures quite very efficiently (but here we are considering the crash stop and network partitioning failures only). Quorum consensus approach is *the best approach* if we talk about *serial isolation level* (i.e. a system that never reads an older version of data) and fault tolerance.

In this paper, we propose a *Circular quorum consensus protocol* for managing the replica servers to achieve *serial isolation level* in distributed transactions. An interesting aspect of this protocol is that for a write operation (*Update transaction*) it may not require to write all copies in the write quorum. Unlike other quorum consensus, it may be as low as a read quorum. Therefore, this protocol increases the write availability and fault tolerance significantly very high. *Circular quorum consensus protocol* uses the *Circular quorum systems* that are very robust in the sense that it not only generalizes the various high-read-capacity quorum systems but also provide the very high availability, fault tolerance and reasonable communication cost. If the reliability of replica sites and intensity of read and write operations vary, then we are able to reconfigure the quorum system of this protocol very smoothly to achieve the desired write availability and fault tolerance.

Organization of this paper is as follows, Next section describes the related work. After that, Section 3 presents the Circular quorum systems. Moreover, Section 4 has the System model for the replicated database system. Furthermore, Section 5 presents the Circular quorum consensus protocol. Section 6 analyzes the Circular quorum systems. While, Section 7 presents a comparative analysis among various replica control protocols those works under an aforementioned scenario. The final Section is a conclusion and future work.

## 2. Related Work

In Majority quorum systems [18] [19], majority of processes give the permission for any read or write operation. This quorum system is quite optimal in terms of the system availability. However, the read capacity is only one as well as read quorum size and read availability is quite not acceptable in most of the applications. Moreover, system load is high and fault tolerance level is quite low. Weighted voting quorum systems [20] generalize the majority quorum systems, such as, if total number of votes in the system are $v$, then $v_r$ votes required for any read operation and $v_w$ votes required for any write operation such that (i) $v_r + v_w > v$ and (ii) $2* v_w > v$. Moreover, we can construct the Majority quorum systems by assigning the equal number of weights to each process. In addition, read capacity, read quorum size and read availability improves by keeping the size of $v_r$ low. However, these improvements come at the cost of the overall system availability and fault tolerance. In this system, System load is also quite very high. If we assign more votes to one process means we are moving toward centralize system rather than distributed system. H. Garcia Molina has contributed in weighted voting systems by explaining how to assign votes in distributed systems [21]. She has introduced the concept of *coteries* and *domination* of the *coteries* in the theory of quorum systems.

FPP quorum systems [22] based on finite projective planes causes the reduction in quorum size up to O ($\sqrt{n}$). The size of quorums in FPP quorum systems is least among fully distributed quorums but the availability is very low in these quorum systems. These quorum systems are optimal in terms of the system load and works well under write dominant scenario. However, the read capacity is only one as well as read quorum size is as high as write quorum size.

Grid quorum systems [22] [23]have a *logical grid* of rows and columns of size *M* and *N,* respectively, a read quorum contains exactly one node from each column while a write quorum contains a read quorum as well as all nodes of any column of the grid. The read capacity of such systems is equal to *M.* In addition, level of fault tolerance and system load is quite better than the Majority quorum systems.

Tree quorum systems [24] organize the system nodes into a complete binary tree of height *h*. We construct a quorum recursively as either (i) the union of the root node and a quorum of any one of the two sub trees or, (ii) the union of the two quorums, one from each sub tree. This system has the read capacity only one and system load is quite very high. In order to reduce the read quorum size and increasing the read capacity, authors again presented the Tree quorum systems [25] [26], that organizes the system nodes into a tree of height *h* structure. Whereas, we construct the read and write quorums recursively as follows, *Read Quorum* = {root} ∨ {majority of read quorums for sub trees of height *h*} and *Write Quorum* = {root} ∪ {majority of write quorums for sub trees of height *h*}. It has the read capacity of order $O(\log n)$ . Smallest quorum size is only one. Nevertheless, this system has major problems as, whenever, the size of the system increases, the read capacity increases very slowly. Moreover, In case of multiple concurrent read operations, most of the read operations work inefficiently as they require more and more replica sites to access. In addition, for a write operation is not possible if the root is inaccessible.  Again, authors have solved the problem of write operations in the case of root inaccessibility by introducing the Generalized-tree quorum systems [27], where we not only take the weighted majority of the children but also take the weighted majority of the levels for read and write quorums.

Hierarchical quorum systems [28] have an n-array tree, where all leaves represent the replica sites. A quorum is formed recursively from the root node, obtaining a quorum in a majority of sub-trees. If we choose the majority for read and write operations as per the aforementioned weighed voting system than we may achieve the slightly higher read capacity. Moreover, it has the minimum write quorum size of $n^{0.63}$.

Wheel quorum systems [29] contain a wheel structure such that it contain $n-1$ *spoke quorums* as {1, $i$} where, $i = 2$ … $n$ and one *rim quorum* as {2 … $n$}. The smallest quorum size in this system is two, and the largest quorum size is $n$ - 1. However, the read capacity is only one and system load is quite very high.

Crumbling walls quorum systems [30] contain $d$ rows and each row may have any number of nodes, Moreover, a quorum consists of one full row plus one representative from every row below the full row. Read and Write quorum size may be as small as one and the write availability of the system is quite high. However, the read capacity of this system is just one and Read quorum size is as high as write quorum size.

Cyclic quorum systems [31], if total number of nodes are $N$, then every cyclic quorum $B_i$ satisfies the following three properties, first, $\forall i : i \in B_i$ where $i \in \{0, 1 \dots N-1\}$, second, $\forall i, \forall j : B_i \cap B_j \neq \varnothing$ where $i, j \in \{0, 1 \dots N-1\}$, and third, $B_i = \{a_1 + i, a_2 + i \dots a_k + i\} \bmod N$, where $a_1$ to $a_k$ all are different integer numbers that lies between 0 to $N$-1. Cyclic quorum system is very close to optimal in terms of quorum size. However, read capacity of these quorum systems are just one and Read quorum size is as high as write quorum size.

Torus quorum systems [32] have the nodes in a rectangular grid such that the last row follows the first row in a circular way, similarly the last column follows the first column in a circular way and we call such a grid as torus. In a torus of $h$ rows, a quorum consists of all the nodes of a row and a representative of each of the $\lfloor h/2 \rfloor$ succeeding rows. In this system, write quorum size is quite smaller than grid quorum systems and the write availability of the system is quite higher than grid quorum systems. However, the read capacity of this system is just one and Read quorum size is as high as write quorum size.

Diamond quorum systems [33] arrange the nodes in the number of rows that forms a two-dimensional diamond like structure, in a general case, there could be any number of rows and each row may have any number of nodes. Moreover, write quorum has all nodes of any one row plus an arbitrary node from remaining rows. Furthermore, read quorum has either any entire row or an arbitrary node from each row. This quorum system achieves a quite high read capacity and low read quorum size. Nevertheless, in case of multiple concurrent read operations while using a diamond structure, most of the read operations work inefficiently as they require more and more replica sites to access.

Symmetric-tree quorum systems [34] join two tree quorum structures together in such a way that leaf nodes are common. We call such trees as up-tree and down-tree. For a read quorum, we chose the root of the either tree. If the root is not available, then all its children are included in the read quorum. Now, each child serves as a new root and process repeats. For a write quorum, we chose the root of the either tree and chose one child of it for the write quorum. Chosen child serves as a new root and process repeats. It improves the read availability over tree quorum systems but it depends on the structure of the tree used.

D-space quorum systems [35] arrange the nodes as similar to the hypercube quorum systems and call it as D-space, In addition, for the construction of a read or write quorum, we divide all dimensions of hypercube into two parts. We choose any single index value from each of the first part dimensions and by having all index values from each of the second part dimensions that construct a subspace $U$ (read-hyper-plane). Similar to that we choose any single index value from each of the second part dimensions and by having all index values from each of the first part dimensions that constructs a subspace $V$ (write-hyper-plane). We can construct a read quorum via having any one read-hyper-plane $U$. and write quorum could be constructed via having a union of one read-hyper-plane $U$ and any one write-hyper-plane $V$. This quorum system is quite suitable for asymmetric access patterns, where read operations are dominant over write operations and provides high read capacity and availability.

Multi-dimensional-grid quorum systems [36] extend the concept of D-space quorum system and may include multiple write-hyper-planes unlike just one in D-space quorum system. Write quorum has the union of any $i$ (out of total $t$) write-hyper-planes and any one read-hyper-plane. In addition, read quorum has a number of nodes of any one read-hyper-plane that lies on any $(t - i + 1)$ write-hyper-planes. This quorum system increases the read availability and read capacity of the D-space quorum system through having multiple write hyper-planes, and provides higher reconfigurable levels of write availability and fault tolerance.

Generalized-grid quorum systems [37], further extends the concept of Multi-dimensional-grid quorum systems. It changes the way through which we select a read quorum. In this quorum system, for a read quorum, we choose any of the $(t - i + 1)$ write-hyper-planes, and then choose one node from each of these write-hyper-planes. It further increases the read availability of the Multi-dimensional grid quorum systems.

## 3. Circular Quorum Systems

In this section, we have presented the Circular quorum systems as follows:

Definition 3.1: *Circular structure C* is a collection of $n$ nodes that we arrange these nodes in a circular way. In addition, every node has a unique number in a sequence, let us say, one to $n$. After that, we divide circular structure $C$ into $k$ arcs, let's say $C_1, C_{2\dots} C_k$ sequentially. Whereas, size of each arc is $n_1, n_2 \dots n_k$ respectively. Moreover, $n_1 + n_2 + \dots + n_k = n$ and $1 \le n_i \le n - k + 1$.

As per the intensity of read/write operations, we can construct the various circular quorum systems by using circular structure $C$. Such as,

## 3.1 Read dominant circular quorum systems

Definition 3.2: *α-circular quorum system* (*α*) is a collection of read quorums and write quorums such that $α = R \cup W$, where, $R$ is a set of read-quorums and $W$ is a set of write-quorums. In addition, we have $\forall i, \forall j$: $r_i \in R$, $w_{i,} \in W$ and $w_j \in W$. Quorum system $α$ must follow the two properties,

$r_i \cap w_j \ne \varnothing$ . **(1)**

$w_i \cap w_j \ne \varnothing$ . **(2)**

For the construction of a write quorum, we choose any $t$ ($1 \le t \le k$) arcs of the C*ircular structure C*, such that, every node is operational among these chosen arcs (let us say, writing arcs. We denote the set of all nodes lies in writing arcs by *WC*). Moreover, we choose any one node from each remaining arcs, i.e.

$w_i = \left\{ x \mid (\forall i, \exists x : (x \in C_i) \wedge (C_i \in C)) \wedge (\forall x : x \in WC) \right\}$. **(3)**

For the construction of a read quorum, either we choose any one node from each of the $(k - t + 1)$ chosen arcs or a complete arc, i.e.

Let, $K = \left\{ 1..k \right\}$. **(4)**

$L = \left\{ l \mid (l \subseteq K) \wedge (\mid l \mid = (k - t + 1)) \right\}$. **(5)**

$r_i = \left\{ x \mid (x \in C) \wedge \left( \left( \forall i, \exists l, \exists x : i \in l, l \in L, x \in C_i \right) \vee (\exists j, \forall x : x \in C_j) \right) \right\}$. **(6)**

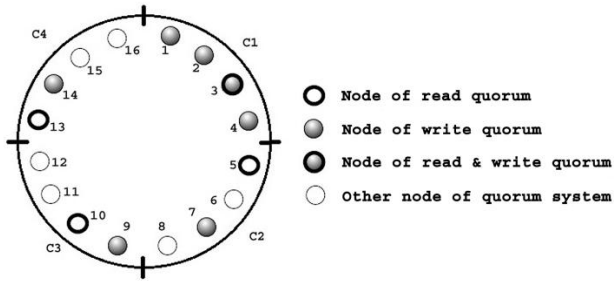

**Fig. 1 α-circular quorum system, where writing arc is only one, i.e. *t* = one**

*Theorem 1:* In α-circular quorum system, any read quorum has non-empty intersection with any write quorum.

*Proof:* here we required to prove that $\forall i$, $\forall j$: $r_i \cap w_j \neq \emptyset$ where $r_i \in R$, $w_j \in W$. Refer to "(6)," read quorum could be constructed either in two ways, by choosing one node from each arc or through a complete arc. If read quorum chooses on node from each arc than Refer to "(3)," write quorum intersects with read quorum because write quorum have at least one writing arc. In another case, if read quorum chooses a complete arc, than Refer to "(3)," write quorum intersect with read quorum. Because, write quorum have at least one node from each arc.

Hence, in *α-circular quorum system* any read quorum has non-empty intersection with any write quorum ∎

*Theorem 2:* In α-circular quorum system, any two write quorums have non-empty intersection.

*Proof:* Here we required to prove that $\forall i$, $\forall j$: $w_i \cap w_j \neq \emptyset$, where, $w_i, w_j \in W$. Refer to "(3)," write quorum $w_i$ have one element from each arc and also have at least one writing arc. Writing arc of $w_i$ will contain the at least one node from any other write quorum $w_j$.

Hence, in *α-circular quorum system,* any two write quorums have non-empty intersection. ∎

*Theorem 3:* Read-one-write-all (ROWA) quorum system is a special case of α-circular quorum system.

*Proof:* Refer to "(3)" and (6)," if *t* is equal to *k* than all nodes of *Circular structure C,* will constitute a write quorum. Furthermore, read quorum will require only one node.

Hence, *Read-one-write-all* (*ROWA*) *quorum system* is a special case of *α-circular quorum system*. ∎

*Theorem 4:* Grid-quorum-system is a special case of α-circular quorum system.

*Proof:* If we represent the columns of the *Grid quorum systems* by arcs and each arc is having equal number of nodes. In addition, if we take *t* = one for *WC*, then Refer to "(3)," write quorum construction of *α-circular quorum system* resembles with the write quorum construction of *Grid quorum systems*.

Moreover, as we have taken $t = 1$, now, for the construction of a read quorum in *α-circular quorum system* we require either to chose one node from each arc or a complete arc. But if we chose only one node from each arc then refer to "(6)," it resembles with the read quorum of *Grid quorum systems*.

Hence, *Grid quorum system* is a special case of *α-circular quorum system*. ∎

*Theorem 5:* Diamond quorum system is a special case of α-circular quorum system.

*Proof:* If we represent the rows of *Diamond quorum systems* by arcs and having $t = $ one for *WC*. Then refer to "(3)" and (6)," read and write quorum construction of *α-circular quorum systems* resembles with the read and write quorum construction of resembles with the *Diamond quorum systems*, e.g. (2, 4, 6, 4, 2) diamond architecture could be achieved by having the arcs C2, C3, C4, C5, C1 as shown in the figure below.

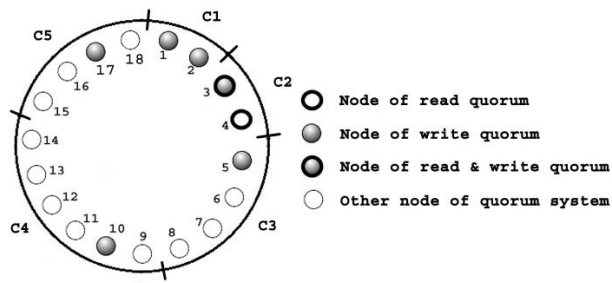

**Fig. 2 α-circular quorum system, which shows that Diamond quorum system, is a special case of α-circular quorum system**

Hence, *Diamond-quorum-system* is a special case of *α circular quorum system*. ∎

*Theorem 6:* D-space quorum system is a special case of α-circular quorum system.

*Proof:* If we represent the write-hyper-planes of the *D-space quorum systems* by arcs that are having equal number of node. In addition, having $t = $ one for *WC* and choosing the $i^{th}$ node of each arc, where, $1 \leq i \leq n/k$ then refer to "(3)," write quorum construction of *α-circular quorum system* resembles with the write quorum construction of *D-space quorum systems*.

Moreover, if we choose $i^{th}$ node of each arc for the construction of a read quorum then refer to "(6)," this read quorum of *α-circular quorum system* resembles with the read quorum of *D-space-quorum systems*.

Hence, *D-space-quorum-system* is a special case of *α-circular quorum system*. ∎

*Theorem 7:* Generalized-grid quorum system is a special case of α-circular quorum system.

*Proof:* If we represent the write-hyper-planes of the *Generalized-grid quorum system* by arcs that are having equal number of node. Then, refer to "(3)," write quorum construction of *α-circular quorum system* resembles with *Generalized-grid quorum system*.

Moreover, if we avoid the complete arc condition for the construction of a read quorum then refer to "(6)," this read quorum of *α-circular quorum system* resembles with the read quorum of *Generalized-grid quorum system*.

Hence, *Generalized-grid quorum system* is a special case of *α-circular quorum system*.  ∎

Definition 3.3: *β-circular quorum system* ($\beta$) is a collection of read quorums and write quorums such that $\beta = R' \cup W'$, where, $R'$ is a set of read quorums and $W'$ is a set of write quorums. In addition, $\forall i, \forall j : r_i' \in R', w_i' \in W'$ and $w_j' \in W'$. Quorum system $\beta$ must follow the two properties,

$$r_i' \bigcap w_j' \neq \varnothing \text{ . (7)}$$

$$w_i' \bigcap w_j' \neq \varnothing \text{ . (8)}$$

For the construction of a write quorum, we choose any $t$ $\left( \lceil (k+1)/2 \rceil \leq t \leq k \right)$ arcs of the *Circular-structure C*, such that, every node is operational among these chosen arcs (let us say, writing arcs. Let denote set of all nodes lies in writing arcs by *WC*), i.e.

$$w_i' = \left\{ x \mid (x \in C) \land (\forall x : x \in WC) \right\}. \text{(9)}$$

For the construction of a read quorum, we choose one node from any $(k\text{-}t+1)$ chosen arcs, i.e.

Let, $K = \left\{ 1..k \right\}$. **(10)**

$$L = \left\{ l \mid (l \subseteq K) \land (\mid l \mid = (k - t + 1)) \right\}. \text{ (11)}$$

$$r_i' = \left\{ x \mid (x \in C) \land \left( \forall i, \exists l, \exists x : i \in l, l \in L, x \in C_i \right) \right\}. \text{ (12)}$$

**Fig. 3 β-circular quorum system, where number of writing arc** $t = \lceil (k+1)/2 \rceil$

*Theorem 8:* In β-circular quorum system, any read quorum has non-empty intersection with any write quorum.

*Proof:* here we required to prove that $r'_i \cap w'_j \neq \emptyset$. Let any read quorum does not intersect with any write quorum. Refer to "(12)," read quorum requires $(k - t + 1)$ arcs. In addition, refer to "(9)," any write quorum requires $t$ arcs. Total read quorum arcs and write quorum arcs should not be greater than $k$, since we have total of $k$ arcs. Let us check, (read quorum arcs + write quorum arcs) = $(k - t + 1) + t = k + 1$. This is a contradiction.

Hence, in *β-circular-quorum-system,* any read quorum has non-empty intersection with any write quorum. ∎

*Theorem 9:* In β-circular-quorum-system, any two write-quorums have non-empty intersection.

*Proof:* here we required to prove that $w'_i \cap w'_j \neq \emptyset$. Let any two write quorums do not intersect with each other. Refer to "(9)," any write quorum requires $t$ arcs. Since, *WC* is having $t$ arcs. Total arcs of two write quorums $(2 * t)$ should not be greater than $k,$ since, we have total of $k$ arcs. Let us check, if we consider the minimum value of $t$ then, $2 * t = 2 * \lceil (k+1)/2 \rceil = (k+2) \text{ or } (k+1)$. This is a contradiction.

Hence, in *β-circular-quorum-system* any two write-quorums have non-empty intersection. ∎

*Theorem 10:* Read-one-write-all (ROWA) quorum system is a special case of β-circular quorum system.

*Proof:* Refer to "(9)" and (12)," if $t$ is equal to $k$ than all nodes of *Circular structure C,* will constitute a write quorum. Furthermore, read quorum will require only one node.

Hence, *Read-one-write-all (ROWA) quorum system* is a special case of *β-circular-quorum-system.* ∎

*Theorem 11:* Majority quorum system is a special case of β-circular quorum system.

*Proof:* Refer to "(9)" and (12)," if we choose $k = n$ then construction of a read and write quorum in *β-circular quorum system* resembles with the construction of a read and write quorum *Majority quorum system.*

Hence, *Majority quorum system* is a special case of *β-circular quorum system.* ∎

## 4.   System Model

A replicated database consist a group of n replica sites, and each replica site contains a unique *replica-id* from the set {1… n}. In addition, any replica site comprises a Transaction manager (**TM**), Scheduler (**SD**), Data manager (**DM**) and a data storage device. These replica sites communicate and coordinate their actions via exchanging messages. Where, message-recipient replica site knows about the address of the source replica site. Moreover, all these replica sites are connected to each other via intranet and/or internet. Furthermore, each replica site contains a copy of the configuration of the circular quorum system.  Any replica site could be either in *working* or *failed* state. **TM** maintains the state of the replica site. In the rest of the paper, we will call a read quorum as a *working read quorum,* if all its replica sites are in *working* stare. Moreover, we will call a write quorum as a *working write quorum* if it contains at least one *working read quorum.*

We assume that any replica site fails in a crash-stop manner and site failures can be detected. Any replica site will be considered as a failure via one of the following reasons, first, it drops the connection from the rest of the system, second, any one of the module of replica site or storage device stops working and third, if it detects that no *working read quorum* exists.

We consider a crash-recovery model where sites can recover and re-join the system after executing a *failure recovery protocol*. In rest of the paper, we will call a replica site as an *active* replica site, if it has a connection with rest of the system as well as all the module of replica site and storage device is in working state. Any *active* replica site may or may not be in a *failed* state depending on the value of state in **TM**.

The database considered here is fully replicated, and thus, each replica site contains a copy of the database. Clients interact with the database by issuing a query or update transaction to one of the *working* replica sites in the system. This replica site (*transaction coordinator*) coordinates its actions with other replica sites (*transaction participants*) in order to execute the query or update transactions.

We assume that all sites receive about the same amount of transactions from the clients. For replicated databases, the correctness criterion is one-copy-serializability. In this criterion, each copy must appear as a single logical copy and the execution of cconcurrent conflicting transactions must be equivalent to a serial execution over all the physical copies. While non-conflicting transactions can be executed concurrently.

## 5.   Circular Quorum Consensus Protocol

In This protocol, every replica site contains the set of data items $D = \{d_{1...}\ d_n\}$. In addition, every data item have a version number to identify the latest update. Moreover, every replica site maintains a timestamp $ts = (C_i,\ i)$ (where $C_i$ is the *Lamport clock* [12] and $i$ is the *replica-id*). Whenever, a client submits the transaction to a replica site $i$, clock $C_i$ increments by one. Furthermore, lets a replica site $i$ sends a message with its clock value $t$ to replica site $j$ then replica site $j$ *updates* the clock $C_j$ by choosing Max $(t + 1, C_j)$. We associate this timestamp with every transaction. Replica site $i$ also maintains the Read-set $R\left(T\left(C_i, i\right)\right)$ (Read-set is a set of all the data items that are to be read by the transaction) and the Write-set $W\left(T\left(C_i, i\right)\right)$ (Write-set is a set of all the data items that are to be written by the transaction).

A transaction $T\ (C_i,\ i)$ is modeled by a tuple $(O_T, <_T)$ where $O_T$ is a set of abstract operations and $<_T$ is a partial order on them. All the Transactions follow the consistency assertions to maintain the consistency. Moreover, individual transactions running in isolation are correct. Even if we operate on database by transactions, there could be two more reasons that could lead the database in inconsistent state, First, multiple replicas and second, concurrent Transactions. Therefore, we require a replica and concurrency control mechanism to keep the database in consistent state. Here we are presenting a more efficient quorum consensus method for replica and concurrency control by using the circular quorum systems that not only provide the consistency among the database but also provide the significantly high fault tolerance and availability.

Transactions can be further classified as query and update transactions. Query transactions have at least one read operation but does not have any write operation, whereas, update transaction have at least one write operation. For query transactions, we have a query workspace over which query transactions operate. In addition, for update transactions, we have a different workspace called update workspace where update transactions operate. We consider here any two transactions, let, $T\ (C_i,\ i)$ and $T\ (C_j,\ j)$ as conflicting if

$$\left(\left(R\big(T(C_i, i)\big) \cap W\big(T(C_j, j)\big) \neq \varnothing\right) \vee \left(R\big(T(C_j, j)\big) \cap W\big(T(C_i, i)\big) \neq \varnothing\right) \vee \left(W\big(T(C_i, i)\big) \cap W\big(T(C_j, j)\big) \neq \varnothing\right)\right).$$

*Transaction coordinator* executes a query transaction by using the *query protocol* as mentioned below.

## 5.1 Query Protocol

**TM** of *transaction coordinator* $i$ multicasts a Read-set $R\left(T\left(C_i, i\right)\right)$ to **TM** of the entire replica sites of a chosen read quorum. After receiving the Read-set from *transaction coordinator* or being a *transaction coordinator*, **TM** sends it to their own **SD** if Commit-set does not have the data items of Read-set $R\big(T\big(C_i,i\big)\big)$. Otherwise, **TM** waits until **TM** fulfills this condition. After receiving the Read-set, **SD** sends this Read-set to the **DM**. Then **DM** places these data items with their version number from permanent memory to the query workspace. Once the data items are present in the query workspace then **SD** reads the data items with their version number of the Read-set $R\big(T\big(C_i,i\big)\big)$ and passes on to the **TM.** After that, **SD** deletes their Read-set $R\big(T\big(C_i,i\big)\big)$. Moreover, **TM** now replies with the value of the data items of Read-set $R\big(T\big(C_i,i\big)\big)$ to the *transaction coordinator's* **TM.** If *transaction coordinator* gets the reply from all requested sites then it chooses the each data item with highest version number. After that, it executes the remaining transaction and sends the desired result to the concerned client. Moreover, after a certain timeout if *transaction coordinator* does not receive a response from all sites of a read quorum then *transaction coordinator* continue to repeat the process by picking another read quorum till the site gets success or if entire read quorums fail to respond. If the entire read quorums fail to respond, then *transaction coordinator* sends a *system-failure* message to the corresponding client and multicasts the *failed* message to the entire system. On receiving the *failed* message, all the modules of the replica site terminate the execution of all transactions. In addition, **TM** sets their state as *failed*. If *transaction coordinator* fails then client issues a fresh query transaction to the any one of the *working* replica site.

*Transaction coordinator* executes an update transaction by using *update protocol* as mentioned below.

## 5.2 Update Protocol

**TM** of *transaction coordinator i* multicasts the transaction T $(C_i, i)$, timestamp $(C_i, i)$, *Read-set $R\big(T\big(C_i,i\big)\big)$ , Write-set $W\big(T\big(C_i,i\big)\big)$*, address of the corresponding client and the set of the chosen write quorum to all other replica sites of a chosen *working write quorum* in a message. On receiving such a message, **TM** places the copy of timestamp $(C_i, i)$ to the all such priority queues where, there exist a timestamp $(C_j, j)$ such that timestamp $(C_j, j)$ is a member of the priority queue and $T$ $(C_i, i)$ conflicts with $T$ $(C_j, j)$. If transaction $T$ $(C_i, i)$ is not in conflict with any other transaction. Then it forms a new priority queue. Moreover, the smallest timestamp in a priority queue has the highest priority in that queue.

Moreover, **TM** sends a *reply T* $(C_i, i)$ message to the *transaction coordinator,* if **TM** holds two conditions, first, timestamp $(C_i, i)$ has the highest priority in all the priority queues where it exists. Second, no transactions $T$ $(C_j, j)$ (where $\forall j$: $i \neq j$, $T (C_i, i)$ and $T (C_j, j)$ are in conflict) is waiting for the response of the *reply T* $(C_j, j)$ message.

Moreover, if **TM** is waiting for the response of a *reply T* $(C_j, j)$ message for the transaction $T (C_j, j)$, which one is not at highest priority now then **TM** sends the *cancel-reply T* $(C_j, j)$ message to the *transaction coordinator*.

Furthermore, throughout the update protocol, if any *transaction participant* awaits a response from the *transaction coordinator* then after a certain timeout, if no response is there, it detects the state of the *transaction coordinator*. If it detects a failure, then it will consider a *working replica site* as a *transaction coordinator* who has the smallest *replica-id* among the *working replica sites* of a chosen write quorum. In addition, it resends the message (for which earlier it was awaiting a response) to the *transaction coordinator*, if it is still a *transaction participant*.

On receiving a *cancel-reply T* $(C_j, j)$ message, **TM** send back the *confirm-cancel-reply T* $(C_j, j)$ message, if it has not received a *reply T* $(C_j, j)$ message from any of the *working transaction participants*. On receiving a *confirm-cancel-reply T* $(C_j, j)$ message, **TM** cancels the reply status for the transaction $T (C_j, j)$.

Furthermore, if *transaction coordinator* receives either a *reply T* $(C_i, i)$ message or detects a site failure from each of the *transaction participants*, then it multicasts a *summit* $(T (C_i, i))$ message to all of the *transaction participants*.

After receiving a *summit* $(T (C_i, i))$ message, **TM** deletes the timestamp $(C_i, i)$ from all priority queues where it exists in **TM**. Moreover, **TM** summits the transaction $T (C_i, i)$ to its **SD** for execution of the transaction.

Once **SD** receives a transaction it places this transaction in a priority queue with the least priority. In addition, it executes all transactions in the same order as present in the queue. **DM** provides the required data items and their version numbers to the update transaction in the update workspace, through asking to the **TM** to launch a *Query Protocol* for the required data items. On receiving the required data items with their version numbers, **DM** places these data items in update workspace for further transaction processing. Once execution of an update transaction $T (C_i, i)$ completes, **SD** launches the *commit protocol*.

## 5.3 Commit Protocol

**SD** sends the *commit-request* $(T (C_i, i))$ message to their **TM**. If the replica site is a *transaction participant* then its **TM** sends the *commit-request* $(T (C_i, i))$ message to the *transaction coordinator*. Moreover, *transaction coordinator* waits until either it receives a *commit-request* $(T (C_i, i))$ message or detects a failure from each *transaction participant*. On receiving the *commit-request* $(T (C_i, i))$ messages from such replica sites that can form a *working read quorum*, *transaction coordinator* multicasts the *commit* $(T (C_i, i))$ message to the chosen write quorum.

After receiving a c*ommit* ($T(C_i, i)$) message, **TM** assigns the Write-set $W\left(T\left(C_i, i\right)\right)$ to the Commit-set i.e. Commit-set $= W\left(T\left(C_i, i\right)\right)$. After that, **TM** passes on this message to **SD** if no read request is pending at **SD**, now **SD** initiates the **DM** to start writing the value of data items, the new version numbers (which we get through incrementing by one in previous version numbers). Once writing completes, **SD** deletes the transaction $T(C_i, i)$ from the queue and intimates to the **TM**. **TM** now assign the empty set to the commit-set i.e. commit-set = Ø. and **TM** sends a *committed* ($T(C_i, i)$) message to the *transaction coordinator*.

On receiving the *committed* ($T(C_i, i)$) messages from such replica sites that can form a *working read quorum*, *transaction coordinator* sends the *committed* ($T(C_i, i)$) message to the client site. On receiving the *committed* ($T(C_i, i)$) message from the *transaction coordinator*, Client site assumes that an update transaction that he has issued is now complete.

On receiving the *committed* ($T(C_i, i)$) messages from such replica sites, that cannot form a *working read quorum*, because of failures of the replica sites. *Transaction coordinator* chooses a *working read quorum* from the entire system. And sends a *restart* message with transaction ($T(C_i, i)$), timestamp ($C_i, i$) and Write-set $W\left(T\left(C_i, i\right)\right)$ to such replica sites of a chosen *working read quorum,* which have not yet committed transaction ($T(C_i, i)$). On receiving such a message, **TM** assigns the Write-set $W\left(T\left(C_i, i\right)\right)$ to the Commit-set i.e. Commit-set $= W\left(T\left(C_i, i\right)\right)$. After that, **TM** sends the *restart* message and transaction $T(C_i, i)$ to its **SD**. Now **SD** stops the execution of the current transaction if it is in execution and reset the update workspace. In addition, it places the transaction $T(C_i, i)$ in priority queue with the highest priority. In addition, **DM** provides the required data items and their version numbers to the update transaction in the update workspace, through asking to the **TM** to launch a *Query Protocol* for the required data items. On receiving the required data items with their version numbers, **DM** places these data items in update workspace for further transaction processing. Now, **SD** starts the execution of the transaction $T(C_i, i)$. After that, on completion of the transaction $T(C_i, i)$, **SD** initiates the **DM** to start writing the value of data items, the new version numbers (which we get through incrementing by one in previous version numbers). Once writing completes, **SD** deletes the transaction $T(C_i, i)$ from the queue and intimates to the **TM**. **TM** now assign the empty set to the commit-set i.e. commit-set = Ø. and **TM** sends a *committed* ($T(C_i, i)$) message to the *transaction coordinator*. Moreover, *transaction coordinator* waits until either it receives a *committed* ($T(C_i, i)$) message or detects a failure from each of the sites to which it has sent the transaction. If *transaction coordinator* does not receives *committed* ($T$ ($C_i$, $i$)) messages from the whole *working read quorum* then *transaction coordinator* repeats the procedure as mentioned in this paragraph, until either it receives *committed* ($T$ ($C_i$, $i$)) messages from the whole *working read quorum*, or it finds no *working read quorum* in the entire system. If no *working read quorum* exist, in the entire system then *transaction coordinator* sends a *failed* message to the entire system. On receiving a *failed* message, all the modules of the replica site terminate the execution of all transactions. In addition, **TM** of replica site sets their state as *failed.*

## 5.4 Failure Recovery Protocol

If any *failed* replica site wishes to re-join the system then its **TM** sets their status as *failed* and deletes all the information that it possesses in their workspace, as well as **SD** and **DM** also deletes the all the information that it possesses in their workspaces.

Moreover, if enough *working* replica sites exist, that can form a read quorum, and then its **TM** sends a *read-all* message to any one of the working replica site from this read quorum. Upon receiving the *read-all* message, **TM** launches the query protocol for all data items and reply with result of the query protocol. Upon receiving the result of the query protocol, **TM** sends this data to the **DM**. and **DM** now places this data into permanent storage. Moreover, now **TM** sets their state as *working.*

Furthermore, because of the massive failures, If not enough *working* replica sites exist, that can form a read quorum; in such a case, entire system reaches into a failed state. In order to start the system, replica site with smallest *replica-id* (we call it *start coordinator*) initiates the system start by multicasting the *prepare-start* message to the entire system. Upon receiving such a message, active replica sites reply with the value and version numbers of all data items. If *start coordinator* does not receive the reply from some of the replica sites then it repeatedly sends *prepare-start* message after certain timeout to such replica sites until it receives the reply from all the replica sites. If *start coordinator* receives the reply from all replica sites then it chooses the highest versions of data items and sends this value to all replica sites. Upon receiving the data, replica site reply with *data-received* message. After receiving the *data-received* message from all replica sites, *start-coordinator* replies with *start* message. Upon receiving a start message, replica site's **TM** sets their state as *working.*

## 6. Analysis of Circular Quorum Systems

Under this section, we are analyzing the various parameters of the Circular quorum systems, such as Quorum size, Fault tolerance, Availability, Read capacity.

## 6.1 Analysis of α-circular quorum system

### 6.1.1 Read quorum size

Refer to "(6)," *Read quorum size* $= \left\{ s \mid \left( s = \left( k - t + 1 \right) \vee \left( \exists i : \left( s = \mid C_i \mid \right) \right) \right) \right\}$.

### 6.1.2 Write-quorum Size

Refer to "(3)," *Write quorum size* $= \left( \left( k - t \right) + \mid WC \mid \right)$.

Where, |WC| = Total number of nodes in any *t* arcs.

### 6.1.3 Fault Tolerance

As per *Circular quorum consensus protocol*, we can execute any query or update transaction even if we have only one *working* read quorum. Therefore, *Fault tolerance* = (*Total number of nodes – Smallest read quorum size*).

*Smallest read quorum size* $= \min_{1 \leq i \leq k} \left\{ \left( \left( k - t + 1 \right), \mid C_i \mid \right) \right\}$.

Hence, *Fault tolerance* $= n - \min_{1 \leq i \leq k} \left\{ \left( \left( k - t + 1 \right), \mid C_i \mid \right) \right\}$.

### 6.1.4 Read Capacity

*Read capacity = Maximum number of disjoint read quorums.*

Refer to "(6)," If we construct the read quorums in *α-circular quorum system* using individual complete arc then total number of disjoint read quorums will be *k*. Moreover, if we construct the read quorums using any one node from each of the (*k − t* + 1) chosen arcs then in order to get the maximum number of disjoint read quorums, we require to choose the available arcs in decreasing order of their size. Furthermore, we can determine the disjoint read quorums out of *x* arcs recursively as follows,

*Disjoint read quorum* (*x*) = *Disjoint read quorum* (*k − t* + 1) + *Disjoint read quorum* (*x* - (*k − t* + 1)), If *x* ≥ (*k − t* + 1).

*Disjoint read quorum* (*x*) = zero, If *x* < (*k − t* + 1).

*Disjoint read quorum* (*k − t* + 1) = Total number of nodes in the smallest arc, out of (*k − t* + 1) chosen arcs.

Hence, *Read capacity* = max {*k, Disjoint read quorum* (*k*)}.

### 6.1.5 System availability

Let $p$ is the probability that a replica server is in a *working* state to perform an operation. In addition, we represent the probability of some $X$ as $P(X)$. Moreover, we represent the availability of the system as $P(S)$. If any read quorum exists in the system then system will be available to perform any transaction.

$P(S) = P(\text{at least one operational read quorum exists})$.

Let, $X$ = at least one operational arc where all the nodes are operational.

$$P(X) = 1 - \prod_{i=1}^{i=k} \left(1 - p^{|C_i|}\right).$$

$P\left(\text{at least one operational node exists in } i^{th} \text{ arc}\right) = 1 - (1-p)^{|C_i|}$.

.Let, $Y$ = at least ($k$-$t$+1) arcs that have at least one operational node in each arc, where $X$ is not true.

"Using (4)," we assume the value of $M$ as $M = \left\{ m \mid \left(m \subseteq K\right) \land \mid m \mid \geq \left(k - t + 1\right)\right\}$.

$$P(Y) = \sum_{\forall m: m \in M} \left( \exists m : \left( \binom{k}{m} \left( \prod_{\forall i: i \in m} \left(1 - (1-p)^{|C_i|} - p^{|C_i|}\right)\right) * \left( \prod_{\forall i: i \in (K \sim m)} (1-p)^{|C_i|}\right)\right)\right).$$

$P(S) = P(X) + P(Y)$.

### . 6.2 Analysis of β-circular quorum system

#### 6.2.1 Read quorum Size

Refer to "(12)," *Read quorum size* $= (k - t + 1)$.

#### 6.2.2 Write quorum Size

Refer to "(9)," *Write quorum size* $= \left(\mid WC \mid\right)$. Where, $\mid WC \mid$ = Total number of nodes in any $t$ arcs.

#### 6.2.3 Fault Tolerance

*Fault tolerance* = (*Total number of nodes − Smallest read quorum size*), whereas, Refer to "(12)," all read quorums are of equal size ($k − t + 1$).

*Smallest read quorum size* $= (k - t + 1)$.

Hence, *Fault tolerance* $= n - (k - t + 1)$.

### 6.2.4 Read Capacity

*Read capacity* = maximum number of disjoint read quorums.

Refer to "(12)," we construct the read quorums in β-*ci*rcular quorum system using any one node from each of the ($k$ - $t$ +1) chosen arcs. In addition, to get the maximum number of disjoint read quorums, it is required to choose the available arcs in decreasing order of their size. Moreover, we can determine the disjoint read quorums out of $x$ arcs recursively as follows,

*Disjoint read quorum* ($x$) = *Disjoint read quorum* ($k - t + 1$) + *Disjoint read quorum* ($x$ - ($k - t$ +1)), If $x \geq (k - t + 1)$.

*Disjoint read quorum* ($x$) = zero, If $x < (k - t + 1)$.

*Disjoint read quorum* ($k$-$t$+1) = total number of nodes in the smallest arc, out of ($k$-$t$+1) chosen arcs.

Hence, *Read capacity = Disjoint read quorum* ($k$)

### 6.2.5 System availability

Let $p$ is the probability that a replica server is in a *working* state to perform an operation. In addition, we represent the probability of some $X$ as $P(X)$. Moreover, we represent the availability of the system as $P(S)$. If any read quorum exists in the system then system will be available to perform any transaction.

$P(S) = P(\text{at least one operational read quorum exists})$.

$P(\text{at least one operational node exists in } i^{th} \text{ arc}) = 1 - (1 - p)^{|C_i|}$

"Using (4)," we assume the value of $M$ as $M = \left\{ m \,|\, \left( m \subseteq K \right) \wedge |\, m |\geq \left( k - t + 1 \right) \right\}$.

Let, $X$ = There are at least ($k - t + 1$) arcs that have at least one operational node in each arc.

$$P(X) = \sum\nolimits_{\forall m: m \in M} \left( \exists m : \left( \binom{k}{m} \left( \prod\nolimits_{\forall i: i \in m} \left( 1 - (1 - p)^{|C_i|} \right) \right) * \left( \prod\nolimits_{\forall i: i \in (K-m)} (1 - p)^{|C_i|} \right) \right) \right).$$

$P(S) = P(X)$.

## 7.   Comparative Analysis

Under this section, we are comparing various suitable (crash stop, static and strict) quorum systems for data replication in the read dominant scenario. It is desirable to have low read quorum size, high read capacity, high fault tolerance, high read availability and quite high system availability. Our proposed Circular quorum consensus protocol under Circular quorum systems achieves all these mentioned properties that we are going to examine here.

## 7.1 Analysis under Read Dominant Scenario

ROWA quorum consensus protocol is very prominent under highly read dominant scenario and provides optimal read capacity, optimal read availability while having no fault tolerance and worst write availability. Moreover, Majority Quorum consensus protocol also provides the optimal read availability at a given read capacity. Nevertheless, under very high read capacity conditions, it provides very low fault tolerance and very low write availability e.g. if we have 1000 nodes, then in order to achieve the read capacity of 500, write quorum size should be of 999 nodes. i.e., system can tolerate just one fault.

Grid quorum consensus, diamond quorum consensus, Generalized-grid quorum consensus are also considered good to achieve high read capacity and high read availability.

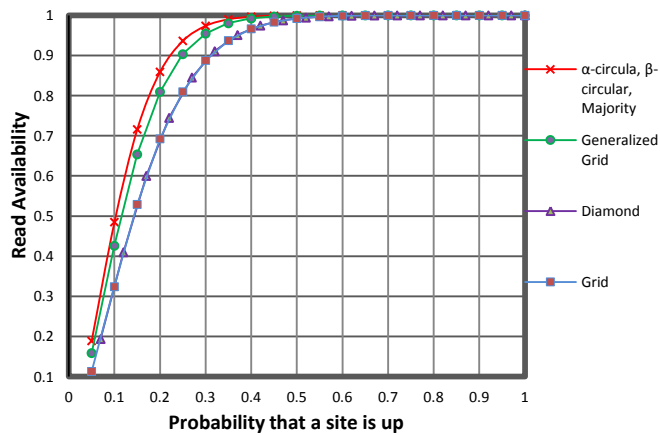

**Fig. 4 Read availability analyses of the various high read-capacity quorum consensus protocols that have 16 replica sites under read capacity of eight**

In order to examine the availability scenario under high read capacity, we are considering here a system of 16 replica sites and desiring a read capacity of eight. Now, we configured all these aforementioned suitable protocols along with our proposed protocol using read dominant α-circular, β-circular quorum systems. In α-circular quorum system we have chosen 8 arcs of 2 replica sites each and t = 7. In β-circular quorum system we have chosen 16 arcs of one replica site each and t = 15. In a generalized grid quorum system, we have chosen a 4 × 4 grid and taken the t = 3. In Grid and Diamond quorum system, we have chosen a 2 × 8 grid.

From fig. 4 it is quite evident that our proposed protocol achieves the optimal read availability as similar to the majority quorum consensus protocol under a given read capacity scenario. However, it occurs at the cost of high write quorum size. Unlike majority quorum consensus, our protocol is so flexible; it may reduce the write quorum size for a given read capacity through compromising the read availability if desired.

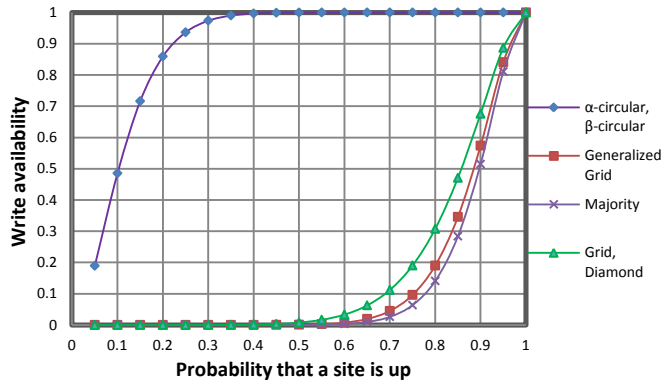

**Fig. 5 Write availability analysis of the various high read-capacity quorum consensus protocols that have 16 replica sites under read capacity of eight**

However, unlike majority quorum consensus, our protocol achieves significantly high write availability as shown in fig. 5. Moreover, α-circular, β-circular quorum systems generalizes the various aforementioned quorum systems. Therefore, these also have the properties of such quorum systems.

## 8. Conclusion

In this paper, we have presented an efficient *Circular quorum consensus protocol* for data replication in order to achieve serial isolation level. Moreover, this protocol handles the transactions by using a suitable Circular quorum system for read dominant scenario. Under read dominant scenario, it provides a significantly high level of fault tolerance, significantly high availability while keeping read capacity very high.

Proposed *Circular quorum systems* generalize the various quorum systems such as Read-one-write-all (*ROWA*) quorum systems, Majority quorum systems, Grid quorum systems, Diamond quorum systems, D-Space quorum systems, Multi-dimensional-grid quorum systems and Generalized-grid quorum systems. Unlike few other quorum systems, this *Circular structure* is so flexible that it can accommodate any number of nodes. Circular quorum systems not only have all the merits of the mentioned quorum systems because these quorum systems are the special cases of the Circular quorum systems, but also have the significant improvement over such quorum systems. In addition, Circular quorum systems are highly reconfigurable, i.e. we can smoothly change the configuration of the Circular quorum systems as per the change in the intensity of the read or write operations. This is very much desirable because the increase in read efficiency comes at the cost of degradation of the write efficiency or vice versa. Moreover, analytically we have proven that Circular quorum systems are quite more efficient among other high-read-capacity strict quorum systems. However, a rigorous simulation study is required to verify these results.